\documentclass[usegraphicx]{mn2e}

\def\etal{{\it et al. }}
\title[Globular Clusters in NGC 2683]
{Keck spectroscopy of globular clusters in the spiral galaxy NGC~2683}

\author[Proctor, Forbes, Brodie \& Strader] {
  Robert N. Proctor$^{1}$\thanks{rproctor@astro.swin.edu.au}, 
   Duncan A. Forbes$^{1}$,
   Jean P. Brodie$^{2}$,
   Jay Strader$^{2}$\\
  $^1$ Centre for Astrophysics \& Supercomputing, Swinburne University, Hawthorn, VIC 3122, Australia\\
  $^2$ Lick Observatory, University of California, Santa Cruz, CA 95064, USA\\
}
\begin{document}
\maketitle  

\begin{abstract}
We analyse Keck spectra of 24 candidate globular clusters (GCs)
associated with the spiral galaxy NGC~2683. We identify 19 bona fide
GCs based on their recession velocities, of which 15 were suitable for
stellar population analysis.  Age and metallicity determinations
reveal old ages in 14 out of 15 GCs.  These old GCs exhibit age and
metallicity distributions similar to that of the Milky Way GC system.
One GC in NGC~2683 was found to exhibit an age of $\sim$3~Gyr. The
age, metallicity and $\alpha$-element abundance of this centrally
located GC are remarkably similar to the values found for the galactic
centre itself, providing further evidence for a recent star formation
event in NGC~2683.

\end{abstract}


\section{Introduction}
\label{intro}

Despite their historical importance in understanding the formation
processes of our own Galaxy (Eggen, Lynden-Bell \& Sandage 1962;
Searle \& Zinn 1978; Mackey \& Gilmore 2004; Forbes, Strader \& Brodie
2004), detailed studies of the stellar populations of globular cluster
(GC) systems in spiral galaxies beyond the Local Group are somewhat
limited.  It is important that this be rectified, not only to inform
formation models of spiral galaxies, but also to constrain formation
models of other morphological types. For example, Ashman \& Zepf
(1992) proposed that the GC systems of elliptical galaxies represent
the merged systems of spiral galaxies plus the addition of newly
formed red (metal-rich) GCs. Bedregal et al. (2006) have argued that
the GC systems of S0s are consistent with faded spirals. A better
understanding of GC systems in spirals, with a range of types and
luminosities, is needed to test these ideas.

Imaging studies of GC systems exist for about a dozen spirals (e.g.
Kissler-Patig et al. 1999; Larsen, Forbes \& Brodie 2001; Goudfrooij
et al. 2003). When sufficient numbers of GCs are present, they reveal
a bimodal colour distribution (similar to those seen in elliptical
galaxies) with the red (metal-rich) subpopulation associated with the
galaxy bulge component (see Forbes, Brodie \& Larsen
2001). Spectroscopic studies of GCs in spirals beyond the Milky Way
and M31 (Burstein et al. 1984; Beasley et al. 2004) are even more
limited. Schroder et al. (2002) investigate the stellar population
properties of 16 individual GCs in M81. Similar, or smaller, numbers
have been investigated in M104 (Larsen et al. 2002), NGC~253 and NGC~300 
(Olsen et al. 2004), and M33 (Chandar et al. 2006). These studies
generally find old ages with a wide range of metallicities for the
GCs. Some GC systems reveal bulk rotation, while others do not, but
small numbers and the lack of edge-on systems make such analyses
uncertain.

\begin{table*}
\begin{centering}
\begin{tabular}[b]{|c|c|c|c|c|c|c|c|c|}  
\hline
Galaxy    & Hubble       &Distance &M$_V$      &Bulge r$_{\rm{eff}}$ & Disc Scale &      Mass HI       &  N$_{GC}$        &  S$_N$   \\
          &    Type      & (Mpc)   &(mag)      & (kpc)          &  (kpc)     &  (10$^9$M$_{\odot}$) &                &          \\
\hline 												    												   
Milky Way & S(B)bc$^1$ &  --     &-20.9$^1$    & 2.5$^3$        &  5.0$^5$        &  4.0$^5$        & 160$\pm$20$^1$ & 0.70     \\
M31       & Sb$^1$     & 0.78    &-21.2$^1$    & 2.4$^3$        &  6.4$^5$        &  3.0$^5$        & 400$\pm$55$^1$ & 1.32     \\  
NGC~2683  & Sb$^2$     & 7.2     &-20.3        & 2.5$^4$        &  1.7$^2$        &  0.6$^2$        & 120$\pm$40$^6$   & 0.90      \\
\hline
\end{tabular}
\caption{Comparison of galaxy parameters for NGC~2683, the Milky Way and M31. 
The values for M$_V$, bulge effective-radius and disc scale-length for NGC~2683
have been 
adjusted, where necessary, to a distance of 7.2~Mpc. The specific frequency 
(S$_N$) is calculated from the number of GCs in the galaxy (N$_{GC}$) 
and the galaxy luminosity (M$_V$) by S$_N$=N$_{GC}$.10$^{0.4(M_V+15)}$. 
References: 1. Courteau \& van den Bergh (1999) and references therein; 2.
Broeils \& van Woerden (1994); 3. van den Bergh (1999); 4. Kent (1985); 
5. Gilmore, King \& van der Kruit (1989); 6. Rhode et al. (2007). 
}
\label{gal_params}
\end{centering}
\end{table*}

Based on the HST Advanced Camera for Surveys (ACS) imaging study of
Forde et al. (2007), we have obtained Keck telescope spectra of GC
candidates in the nearby, edge-on Sb spiral NGC~2683. A variety of
distance estimates exist in the literature for NGC~2683. Here we adopt
the surface brightness fluctuation distance modulus of Tonry et al.
(2001) modified by the correction found by Jensen et al. (2003). This
is gives m--M = 29.28 $\pm$ 0.36 or 7.2 $\pm$ 1.3 Mpc which lies near
the midpoint of the literature estimates. With a luminosity of M$_V$ =
--20.31~mag it has a lower luminosity (by a factor of 2) than the
Milky Way or M31.  We note that, although possessing a Hubble type and
bulge size similar to the Milky Way and M31, NGC~2683 exhibits a disc
size and HI gas mass that are significantly smaller.  Rhode et
al. (2007) show that the extent of the GC system of NGC~2683 is also
rather small, with the projected density of the system falling to
background levels within $\sim$8~kpc. This can be compared to the
Milky Way GC system in which a fraction of the GC system
lies outside $\sim$30~kpc.  Some properties of NGC~2683 are compared
to those of the Milky Way and M31 in Table \ref{gal_params}.

In Section \ref{obs} we present our observations and data reduction
methods.  The measurement and analysis of recession velocities and
Lick indices are given in Section \ref{spec_anal}. The results of our
chemical and kinematic analysis of the sample is outlined in Section
\ref{results}.  Our conclusions are presented in in Section \ref{disc}.

\section{Observations and data reductions}
\label{obs}

Spectra of 24 GC candidates around NGC~2683 were obtained with the Low
Resolution Imaging Spectrometer (LRIS; Oke \etal 1995) on the Keck I
telescope. Candidate selection, based on ACS imaging data, is detailed
in Forde et al. (2007). Briefly, our spectroscopic sample was selected
from amongst the brightest of GC candidates. The candidates were
chosen to represent both red and blue subpopulations. While not a full
statistical sample, the candidates are therefore 
representative of the GC system as a whole. It should also be noted that,
with GCs only partially resolved in the HST imaging, it is to be
expected that the sample will include some foreground stars.

Spectral observations were obtained in 2005 February 07--08 with an
integration time of $16 \times 1800$s = 8 hours.  Seeing was $\sim$~1
arcsec on both nights.  A 600 lines-per-mm grating blazed at 4000 \AA\
was used on the blue side, resulting in a wavelength range of 3300 --
5900 \AA\ and a FWHM spectral resolution of $\sim3.3$ \AA. The spectra
were not flux calibrated.

\begin{figure}
\includegraphics[width=9cm,angle=-90]{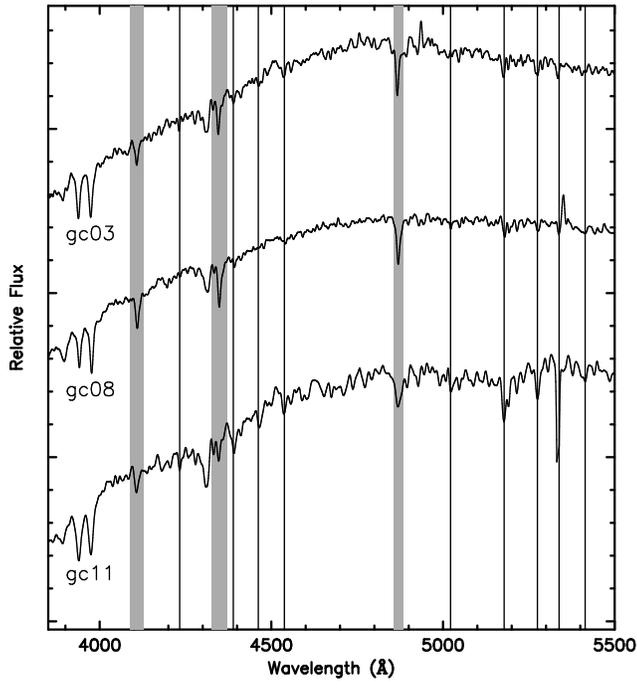}
\caption{A sample of GC spectra after broadening to the Lick resolution. The 
age sensitive Balmer lines are highlighted in grey. Metallicity sensitive 
features are marked by lines. The young GC (gc11; bottom) is compared to old 
GCs with similar colours (gc03 and gc08). Candidate gc11 has similar Balmer 
line-strengths, but significantly stronger metallicity-sensitive features than 
gc03 and gc08, indicating a younger, more metal-rich stellar population.} 
\label{spectra}
\end{figure}

Data reduction was carried out using standard IRAF\footnote{IRAF is
distributed by the National Optical Astronomy Observatories, which are
operated by the Association of Universities for Research in Astronomy,
Inc., under cooperative agreement with the National Science
Foundation} commands.  The tracing of spectra and
background-subtraction was done using the command {\it apall}.
Comparison lamp spectra were used for wavelength calibration (mostly
based on 8 Hg lines). Zero--point corrections of up to 1.5~\AA\ were
performed on the science spectra using the bright [OI] skyline at 5577
\AA. The 16 individual spectra of each GC candidate were then
average-combined with 3-$\sigma$ clipping.  A sample of the GC spectra
are shown in Fig. \ref{spectra}.  The backgrounds subtracted from the
spectra of the GC candidates -- admixtures of background galaxy light
and sky -- were retained for analysis of the galaxy rotation
curve. These were average combined and then sky-subtracted. The
removal of the sky was achieved by identifying the GC candidate with
the lowest background level; gc01 -- a candidate lying at large radial and
azimuthal distances from the galaxy centre (see Fig. \ref{position};
right). The candidate was also found to have signal-to-noise
$<$1~\AA$^{-1}$, and therefore probably contained no object.  The
background of this candidate, which lies in the halo of the galaxy, is
therefore the least contaminated by either galaxy or globular cluster
light. Indeed, cross-correlation of the spectrum of the background of
this candidate with the solar spectrum gave a recession velocity less
than 1~km s$^{-1}$, indicating very little contamination from
NGC~2683. By scaling this spectrum by the total flux in the bright OI
skyline at 5577 \AA, estimates of the sky levels in the backgrounds of
other candidates could be made and subtracted.  The residuals from
this process are therefore estimates of the spectra of the background
galaxy. These were used to measure a rotation curve for the galaxy,
but were not subject to stellar population analysis. 

We also measured the gas kinematics in NGC~2683 using the bright
[OIII]$\lambda$5007~\AA\ emission-lines evident in most of the galaxy
spectra.  We have therefore been able to measure GC, stellar
\emph{and} gaseous recession velocities from the majority of slitlets
in the mask.

\begin{figure}
\includegraphics[width=8cm,angle=0]{comb_position.ps}
\caption{The distribution of GC candidates with respect to the galaxy
NGC~2683 is shown. The red diamond marks the centre of the galaxy,
while the line identifies the major axis. Red stars are
candidates identified as being stars. Blue stars represent candidates
contaminated by OB associations. Filled symbols are GCs for which
stellar population analysis was performed.  Open circles are
candidates whose signal-to-noise was too low for stellar population
analysis. The open circle on the extreme right is the candidate used
for sky estimation (Section \ref{obs}).}
\label{position}
\end{figure}

\begin{figure}
\includegraphics[width=9cm,angle=0]{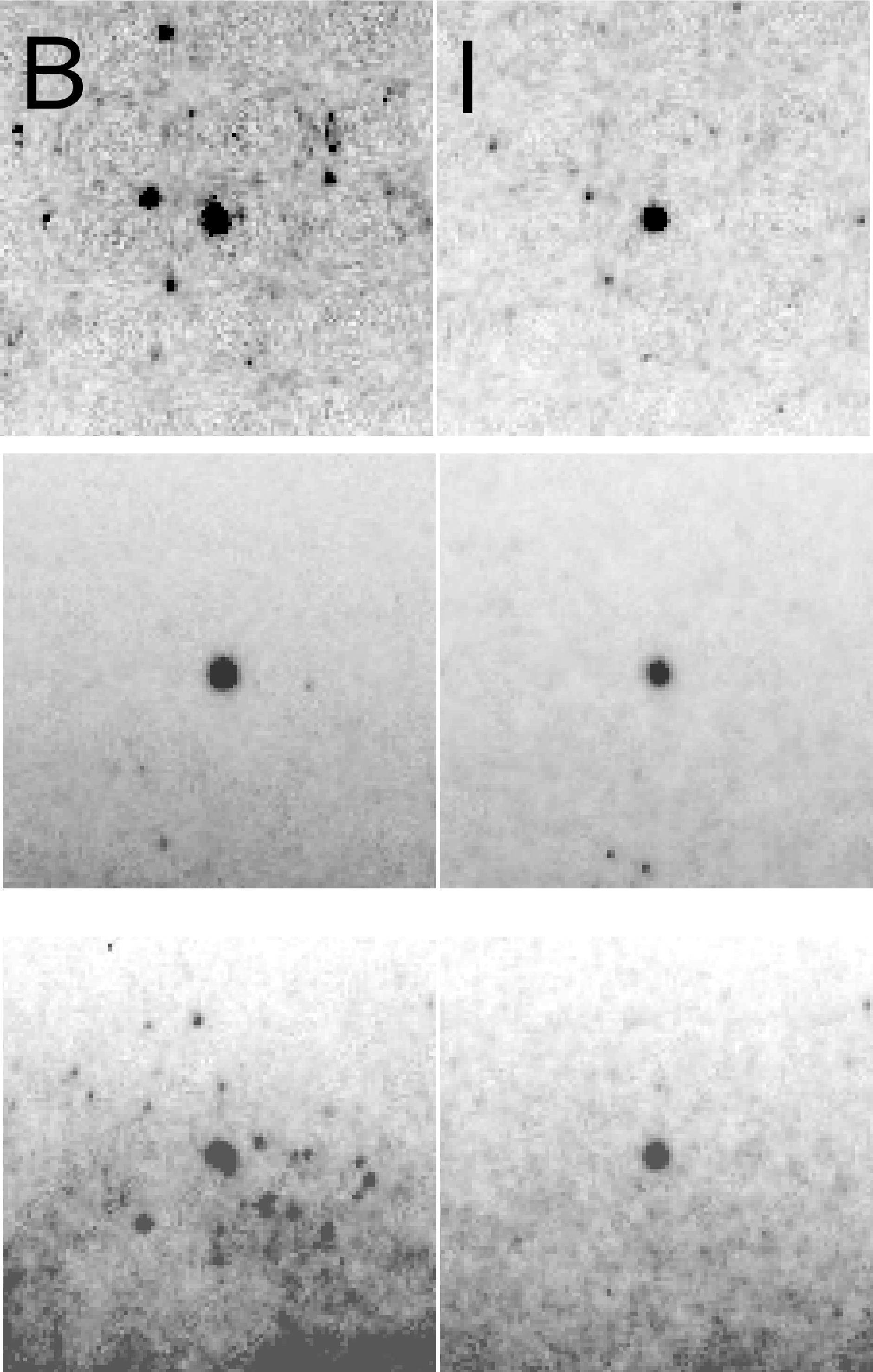}
\caption{ACS imaging showing three GC candidates (from top to bottom;
gc05, gc11 and gc12). Each image (B band; left and I band; right) is
10 arcsec ($\sim$360~pc at 7.2~Mpc) on a side.  Candidates gc05 and
gc12 are clearly projected against OB associations in NGC~2683 itself,
and contamination appears highly likely. On the other hand, there is no
evidence for an OB association affecting the young GC candidate;
gc11.}
\label{gc12}
\end{figure}

During the reductions, three GC candidates (gc02, gc15 and gc24) were
identified by their spectra to be stars. Four others (gc01,
gc13, gc20 and gc23) were found to have signal-to-noise ratios below
10~$\AA^{-1}$. A visual inspection
of the Forde et al. (2007) imaging identified two other candidates
to be contaminated by stars in OB associations in NGC~2683 itself (see
Fig. \ref{gc12}). All nine of the candidates identified above were
therefore excluded from our stellar population analysis. The sample
therefore contains 19 GCs suitable for recession velocity (RV) analysis and
15 GCs suitable for stellar population analysis. Details are given as
Notes in Table \ref{gc_params}.

\begin{table*}
\begin{centering}
\begin{tabular}[b]{c|c|c|c|c|c|c|c|c|c|c|c|c|}  
\hline
ID&     RA     &  Dec   &Note   &       RV$_{GC}$ & RV$_{Gal}$ &RV$_{Gas}$  &     r	 &   z      &    B$_0$     &     I$_0$   &   (B-I)$_0$   &  Size\\
&  (J2000)   & (J2000)&       &        (km s$^{-1}$) & (km s$^{-1}$)   &  (km s$^{-1}$)  & (``)    &   (``)    &  (mag)   &   (mag) &  (mag)   &  (pc)    \\
		      \hline 	 
gc01 & 	8:52:27.8& 33:24:40.4&S/N$<$1 (Sky)  &      --        &    --    &    --   &     --  &    --   &      --     &    --    &     --    &   --    \\
gc02 & 	8:52:27.8& 33:24:40.4&Star & 	  --        &    --    &    --   &   147.4 &   96.3  &     20.97   &    19.10 &    1.87   &   1.24    \\
gc03 & 	8:52:37.9& 33:23:24.4& --  & 	368(3)      & 	559(7) & 525(20) &   111.7 &  -46.9  &     21.12   &    19.52 &    1.60   &   1.92     \\
gc04 & 	8:52:35.3& 33:24:31.4& --  & 	602(16)     & 	552(1) & 590(20) &    87.3 &   23.5  &     22.87   &   20.86  &    2.02   &   2.27    \\
gco5 & 	8:52:35.5& 33:24:39.7&Contam & 	283(14)     & 	546(2) & 593(20) &    79.7 &   27.6  &     23.28   &   21.39  &    1.89   &   2.66     \\
gc06 & 	8:52:38.2& 33:24:07.5& --  & 	491(3)      & 	563(2) & 530(20) &    78.6 &  -19.0  &     22.14   &   20.44  &    1.70   &   2.31     \\
gc07 & 	8:52:35.3& 33:25:22.5& --  & 	557(14)     & 	532(4) & 497(20) &    59.6 &   51.2  &     22.83   &   21.33  &    1.51   &   2.86    \\
gc08 & 	8:52:35.6& 33:25:35.5& --  & 	560(4)      & 	396(5) & 437(20) &    66.2 &   39.3  &     21.24   &   19.72  &    1.52   &   4.17    \\
gc09 & 	8:52:41.0& 33:24:51.7& --  & 	388(4)      & 	481(2) & 471(20) &    22.5 &  -12.6  &     20.95   &   18.98  &    1.97   &   2.96     \\
gc10 & 	8:52:43.5& 33:24:16.8& --  & 	317(15)     & 	447(31)& 477(20) &    25.0 &  -59.4  &     22.84   &   20.86  &    1.98   &   2.67    \\
gc11 &  8:52:42.0& 33:25:02.9& --  &    430(3)      & 	461(8) & 460(20) &     5.7 &  -13.5  &     21.33   &   19.82  &    1.51   &   5.55     \\
gc12 &  8:52:42.3& 33:25:11.3&Contam&   432(3)      & 	432(3) & 424(20) &    -2.9 &  -10.2  &     21.88   &   20.00  &    1.89   &   3.09     \\
gc13 &  8:52:43.6& 33:24:59.4&S/N$<$10 &  --         & 	500(27)&  --     &    -6.0 &  -30.2  &     22.77   &   20.89  &    1.88   &   2.04     \\
gc14 & 	8:52:44.4& 33:25:18.6& --  & 	304(5)      & 	426(2) & 424(20) &   -26.6 &  -23.7  &     21.27   &   19.43  &    1.84   &   2.22    \\
gc15 & 	8:52:42.1& 33:26:30.8&Star & 	 --         & 	351(5) & 386(20) &   -57.3 &   47.8  &     20.86   &   18.47  &    2.39   &   1.01      \\
gc16 & 	8:52:42.8& 33:26:45.3& --  & 	217(5)      & 	344(5) &  --     &   -73.8 &   51.8  &     22.05   &   19.97  &    2.08   &   2.96     \\
gc17 &  8:52:43.1& 33:26:54.1& --  &    435(5)      & 	385(7) &  --     &   -82.7 &   55.4  &     21.57   &   19.84  &    1.73   &   9.66   \\
gc18 & 	8:52:46.7& 33:25:52.9& --  & 	244(6)      & 	365(2) &  --     &   -71.2 &  -19.8  &     22.48   &   20.99  &    1.49   &   2.18     \\
gc19 & 	8:52:47.7& 33:25:54.1& --  & 	352(4)      & 	371(4) & 304(20) &   -80.9 &  -27.8  &     20.85   &   19.20  &    1.65   &   1.89    \\
gc20 &  8:52:48.0& 33:26:26.1&S/N$<$10 & 247(21)     & 	295(2) & 237(20) &  -106.2 &   -7.8  &     23.61   &   21.62  &    1.99   &   1.77     \\
gc21 & 	8:52:50.4& 33:25:46.9& --  & 	711(5)      & 	259(5) & 329(20) &   -99.8 &  -56.8  &     21.89   &   20.50  &    1.40   &   2.52     \\
gc22 & 	8:52:49.0& 33:26:30.9& --  & 	307(49)     & 	307(2) & 329(20) &  -118.5 &  -13.3  &     23.01   &   21.54  &    1.47   &   2.67     \\
gc23 &  8:52:50.0& 33:26:45.9&S/N$<$10 & 456(43)     & 	285(4) &   --    &  -137.9 &  -11.5  &     23.35   &   21.38  &    1.97   &   3.07    \\
gc24 & 	8:52:53.0& 33:27:15.3&Star & 	 --        &   166(85)& 252(20)  &  -185.2 &  -17.3  &     22.09   &   20.58  &    1.52   &   1.31   \\
\hline
\end{tabular}
\caption{Key values for the GC candidates. Recession velocities for GC
candidates, background galaxy stars and gas are derived from
our spectral analysis. Coordinates, Galactic extinction corrected
photometric properties and candidate sizes are from the HST study of
Forde et al. (2007). Radial distance along the major axis (r) and
vertical distance along the minor axis (z) were derived from these HST
data. The notes denote our classification of the candidate
(see Section \ref{meas}).}
\label{gc_params}
\end{centering}
\end{table*}

\begin{table}
\begin{centering}
\begin{tabular}[b]{|c|c|c|}
\hline
Index & Offset & Error \\
\hline
H$\delta_A$   &     0.373&   0.254\\
H$\delta_F$   &     0.007&   0.127\\
CN$_1$       &    -0.001&   0.011\\
CN$_2$       &     0.005&   0.012\\
Ca4227       &     0.298&   0.091\\
G4300        &     0.142&   0.211\\
H$\gamma_A$   &    -0.460&   0.170\\
H$\gamma_F$   &     0.011&   0.059\\
Fe4383       &   0.583  & 0.188\\
Ca4455       &   0.420  & 0.091\\
Fe4531       &   0.180  & 0.108\\
C4668        &  -0.846  & 0.073\\
H$\beta$       &   0.032  & 0.123\\
Fe5015       &   0.696  & 0.140\\
Mg$_1$       &     0.027&   0.005\\
Mg$_2$       &     0.052&   0.003\\
Mgb          & -0.115   &0.050\\
Fe5270       &   0.086  & 0.083\\
Fe5335       &   0.384  & 0.114\\
Fe5406       &  -0.025  & 0.068\\
\hline
\end{tabular}
\caption{Index offsets required to match the Lick system (Section \ref{meas}). Errors are taken as the error on the mean (i.e. rms/$\sqrt{6}$).}
\label{c2l}
\end{centering}
\end{table}

\section{Spectral analysis}
\label{spec_anal}
In the following we outline the spectral analysis from which we
measure recession velocities and stellar population properties. We
also make use of the HST photometric measurements of Forde et
al. (2007) (see Table \ref{gc_params}).

\subsection{Kinematics}
\label{kins}
The recession velocities of GC candidates and their galaxy
backgrounds were determined by cross-correlation against six high
signal-to-noise stellar templates using the IRAF command
\emph{fxcor}. The heliocentric velocities of the templates themselves
were measured by cross-correlation against a high resolution solar
spectrum. The average of the values of RV derived from comparison to
the six template stars was taken as the measured value, while the rms
scatter was taken as the error.  The RVs of the galaxy's gas were also
measured in the background galaxy spectra. This was achieved by the
fitting of Gaussians to the bright [OIII]$\lambda$5007 emission lines
evident in most galaxy spectra.

The results of this analysis are presented in Section \ref{results}.
We next detail the determination of the properties of 
the stellar populations in our sample of GCs using Lick indices.

\subsection{Measurement and analysis of Lick indices}
\label{meas}
We measured Lick indices using the definitions of Trager et al. (1998)
and Worthey \& Ottaviani (1997). Indices were measured after
convolving the spectra with the Gaussians required to broaden to the
wavelength-dependent Lick resolution (Worthey \& Ottaviani 1997).
Lick indices and their associated errors are shown in Table
\ref{indices}.  Calibration to the Lick system was performed using 6
Lick standard stars. The additive corrections required to match the
Lick system and their errors are given in Table \ref{c2l}.

The measured indices were then compared with SSP models. We elected to use the 
recent models of Lee \& Worthey (2005) combined with Houdashelt (2002) 
sensitivities to abundances ratios. We detail the method by which the 
SSP models are combined with the Houdashelt sensitivities in  Mendel et al. 
(2007), in which we show that this 
combination reproduces the ages, metallicities and 
`$\alpha$'--abundance ratios of Galactic globular clusters extremely well. 
This gives us confidence in making direct comparisons of our results with 
the Galactic globular cluster system (Section \ref{results}).

The comparisons to SSP models were carried out using the $\chi^2$-fitting 
procedure of Proctor \& Sansom (2002)
(see also Proctor et al. 2004a,b and Proctor et al. 2005) to measure the 
derived parameters; log(age), [Fe/H], [Z/H] and [E/Fe] (a proxy for 
the `$\alpha$'--abundance ratio; see Thomas et al. 2003 for details).
Briefly, the technique for deriving these parameters
involves the simultaneous comparison of as many observed indices as possible 
to models of single stellar populations (SSPs). The best fit is found by 
minimising the deviations between 
observations and models in terms of the observational errors, i.e. $\chi$.
We have shown this approach to be relatively robust
with respect to many problems which are commonly experienced in the
measurement of spectral indices and their errors. These include poor or no 
flux calibration, poor sky subtraction and poor calibration to the Lick system.
The method is similarly robust with respect to many of the uncertainties in 
the SSP models used in 
interpretation of the measured indices; e.g. the second parameter effect 
in horizontal branch morphologies and the uncertainties associated with the 
Asymptotic-Giant Branch. It was shown in Proctor et al. 
(2004a) and Proctor et al. (2005) that the results derived using the $\chi^2$ 
technique are, indeed, significantly more reliable than those based on only a 
few indices. 

\begin{figure}
\centerline{\includegraphics[width=9cm,angle=-90]{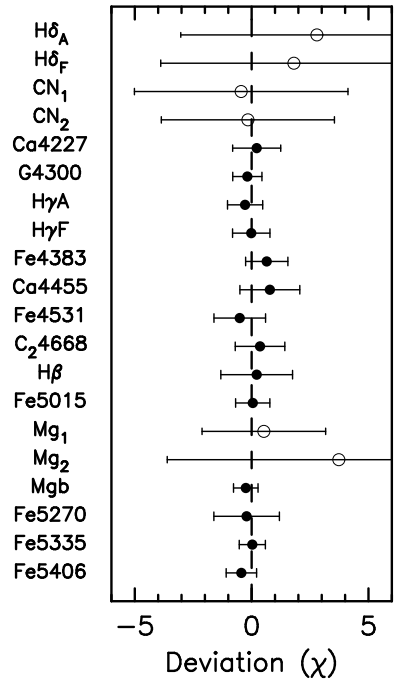}}
\caption{The average deviations from the best fits are shown in terms of 
observational errors (i.e. $\chi$) . The open symbols identify the indices 
excluded from the fitting procedure. Error bars show rms scatter.}
\label{chis}
\end{figure}

The process by which the candidate spectra were compared to the models
was iterative. First, fits were obtained for all the candidates using
all the available indices. The patterns of deviations from the fits
obtained was then used to identify individual indices that matched the
models poorly (see Fig.  \ref{chis}). These included the H$\delta$, CN
indices for which flux levels were generally too low for accurate
determination and Mg$_1$ and Mg$_2$ indices which suffer from flux
calibration sensitivity. These indices were excluded from the analysis
and the fits performed again. These fits were carried out using a
clipping procedure in which indices deviating from the model fit by
more than 3$\sigma$ were excluded, and the fit performed again. Many
of these poorly fitting indices could be associated with known
problems, e.g. the contamination of the Mgb index by the 5202~\AA\
sky-line in low signal-to-noise candidates.  Indices that are excluded
on this basis are in parentheses in Table \ref{indices}.  On average,
after all exclusions, 10 indices were used in each of the final fits.

For each GC in the sample, errors in the derived parameters (log(age),
[Fe/H], [E/Fe] and [Z/H]) were estimated using 50 Monte-Carlo
realisations. Best-fit model indices were perturbed by Gaussians, the
width of which were set equal to the observational errors added in
quadrature to the errors in offset to the Lick system (Table
\ref{c2l}). Error estimates in the derived parameters are therefore
highly sensitive to the estimates of index errors. The process also
makes no allowance for the \emph{correlated} components of the
observational index errors, such as velocity dispersion, flux
calibration and background subtraction errors.  The errors are
modelled instead as purely random Gaussian distributions. As a
consequence, our error estimates must be considered to include both
random \emph{and} systematic errors.

\section{Results}
\label{results}
\subsection{Results of recession velocity analysis}
\label{rv_res}

\begin{figure}
\centerline{\includegraphics[width=8cm,angle=-90]{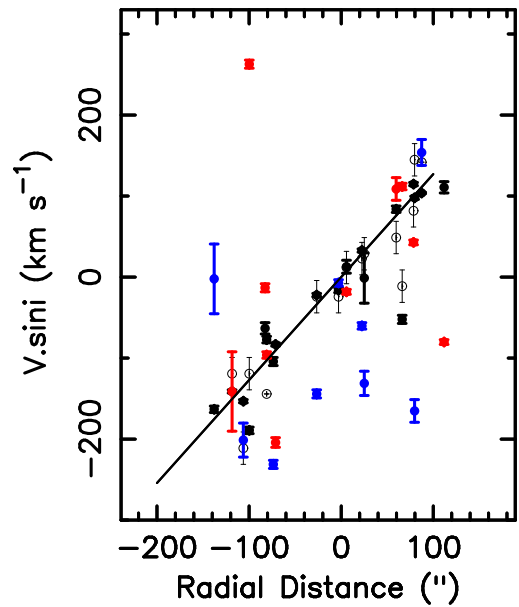}}
\caption{ Recession velocities are shown against radial distance from
the galactic centre perpendicular to the minor axis. Solid black
points are stellar data, while open symbols for the gas. The stellar and
gas data clearly show rotation.  The least-squares fit to the stellar
data is shown as a line.  The small scatter about the fit shows the
rotation to be roughly cylindrical.  Coloured points are the GC data
with colour representing actual GC colour from Forde et al. (2007) for
red (B-I$>$1.8) and blue (B-I$<$1.8) candidates.}
\label{comb_galrot}
\end{figure}

The results of our analysis of recession velocities are given in Table
\ref{gc_params} and are presented in Fig. \ref{comb_galrot}.  
The value assumed for the galactic centre (442.8 km s$^{-1}$) was
estimated such that the value of the least-squares fit to the stellar
RVs with radial distance (solid line in 
Fig. \ref{comb_galrot}) passes through 0.0 km s$^{-1}$ at a radial
distance of 0.0 arcsec. Note that candidates identified as stars and
the single object with signal-to-noise $<$1 are omitted from these and
all subsequent plots of GC data. We therefore present recession
velocities for 19 GCs.

The stellar and gas emission-line data (Fig. \ref{comb_galrot}) clearly show an
increasing rotation speed with increasing radial distance from the
galactic centre, and can be seen to be essentially cylindrical
(i.e. there is little scatter and no particular trend in RV with
distance above or below the least-square fit). The figure also shows
the rotation of gas and stars to be in very good agreement.

It is evident that we do not reach the radii at which rotation
is observed to flatten. However, our results are nevertheless
consistent with the rotation of Casertano \& van Gorkom (1991)
and Broeils \& van Woerden (1994), who find similar rotation curves
with a flattening/peak lieing just beyond the range probed by our
data. The apparent dip in stellar and gaseous RVs at radial distance
65 arcsec in the rotation profile of Barbon \& Capaccioli (1975) is
also present at similar radii in our data (Fig. \ref{comb_galrot};
lower left), although we note that they find a significantly steeper
rotation curve than Casertano \& van Gorkom (1991), Broeils \& van
Woerden (1994) or ourselves. Our `dip' is also significantly deeper
than that observed by Casertano \& van Gorkom, with both stars and gas
rotating in the opposite sense to the rest of the galaxy at similar
radii.

The RVs measured in the GC spectra are also shown in 
Fig. \ref{comb_galrot}. GC recession velocities are generally
consistent with those of the stars and gas. However, we lack sufficient
numbers to unambiguously identify rotation in the GC system.\\

\subsection{Results of stellar population analysis}
\begin{figure}
\includegraphics[width=8cm,angle=0]{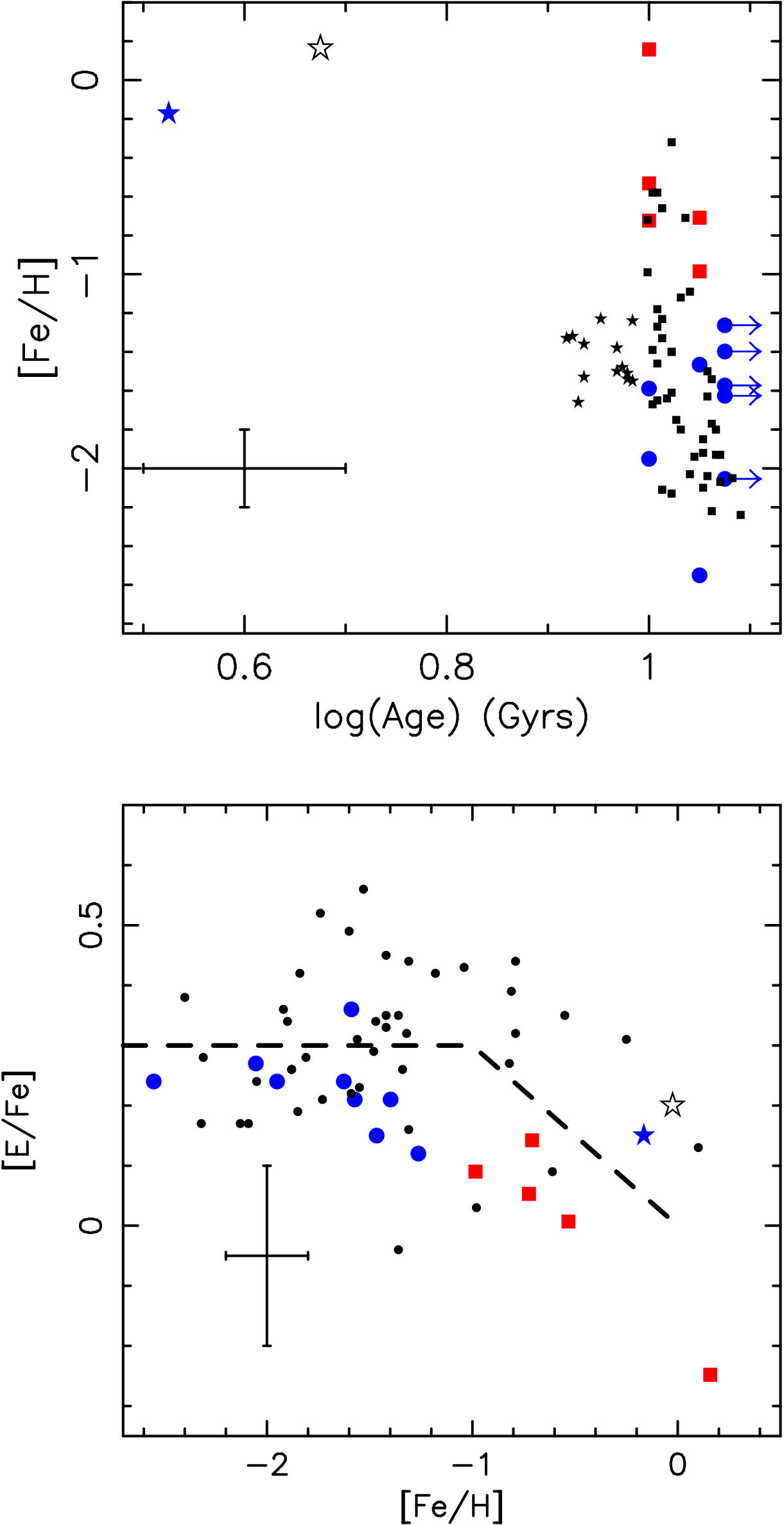}
\caption{Metallicity--age (top) and $\alpha$-element
abundance--metallicity (bottom) relations for NGC~2683 GCs.  Values
for the Galactic GC system are shown as small solid symbols. These are
from de Angeli et al. (2005) (top) and Pritzl et al. (2005)
(bottom). The relation for
local Galactic stars is shown as a dashed line in the bottom plot for
reference only. GCs in NGC~2683 with (B-I)$_0>1.8$ are shown as red
squares, those with (B-I)$_0\leq1.8$ as blue circles. Arrows indicate
GCs whose derived age equals the maximum modelled by Lee \& Worthey
(2005). One GC (gc11; solid blue star) exhibits age, [Fe/H] and [E/Fe]
similar to the those measured in the galactic centre by Proctor \&
Sansom (2002) (open black stars). Combined systematic and random errors
from our Monte Carlo analysis (Section \ref{meas}) are indicated in
each plot.}
\label{agez}
\end{figure}

The results of our age and metallicity determinations are given in
Table \ref{tab_agez} and plotted in Fig.  \ref{agez}.  Candidates with
signal-to-noise $<$10 are excluded from our analysis, leaving 15 GCs
suitable for stellar population analysis.

In Fig. \ref{agez} our results are compared to the values for Galactic
GCs from de Angeli et al. (2005) and  Pritzl, Venn \& Irwin (2005).  It is
shown in Mendel et al. (2007) that ages and metallicities derived from
Lee \& Worthey (2005) SSP models agree extremely well with Galactic GC
measurements from colour-magnitude diagrams and high resolution
spectral studies. For [Fe/H], Mendel et al. find only a 0.028$\pm$0.024~dex
average offset between the value derived from Lee \& Worthey SSPs
models and the Harris (1996) values for 42 Galactic GCs. A similar
offset ($\sim$--0.024$\pm$0.021~dex or --0.28$\pm$0.24~Gyr) was found
in the comparison of the derived ages with the data from de Angeli et
al. (2005). Finally, the average values of [E/Fe] derived by Mendel et
al. (2007) for Galactic GCs are offset from the Pritzl et al. (2005)
values by --0.024$\pm$0.02~dex (T. Mendel 2007; private communication). The
Mendel et al. (2007) results are consequently fully consistent with the
literature data. We therefore have reasonable confidence in our
comparison of the ages and metallicities of GCs of NGC~2683 with those
of the Milky Way.\\

\subsection{Consistency checking}
\begin{figure}
\includegraphics[width=8cm,angle=-90]{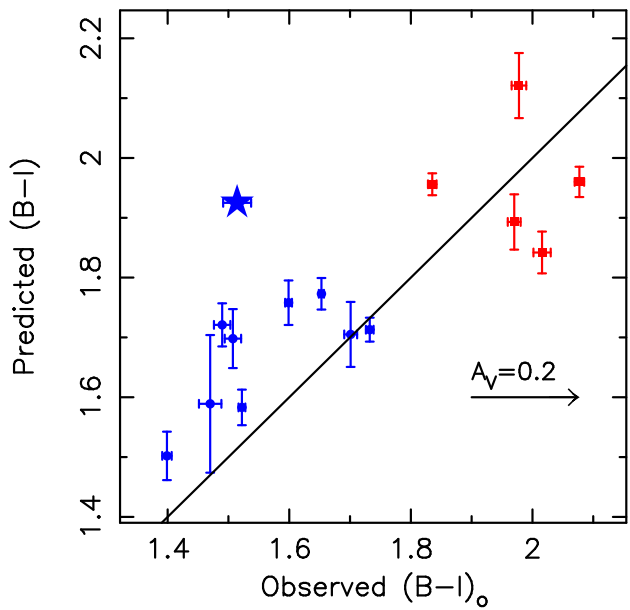}
\caption{A comparison between (B-I) values predicted from BC03 models
using our derived ages and metallicities and the observed Galactic
extinction corrected colours of Forde et al. (2007). Symbols as
Fig. \ref{agez}.  The solid line represents the one-to-one
relation. The extinction correction corresponding to A$_V$=0.2 is
shown in the bottom right.}
\label{colour_comp}
\end{figure}

However, before interpreting the ages and metallicity estimates, we
sought to gain further confidence in our results by using them to
predict the B-I colours of our GC sample (using the SSP models of
Bruzual \& Charlot 2003) for comparison to the observed HST colours
(Forde et al. 2007). The comparison is shown in
Fig. \ref{colour_comp}. The predictions compare quite favourably with
the observed values, particularly given the $\sim$0.15~mag
overestimation of predicted (B--I) colour found by Pierce et al. (2005, 2006)
in similar studies. This is believed to be primarily the effect of the
poor modelling of the horizontal branch (see also Strader \& Smith
2007). Scatter should also be expected to be relatively high in our
study due to the highly variable internal extinction in NGC~2683.

There is, however, one clearly aberrant GC -- gc11; an apparently
young GC (Table \ref{tab_agez}). We note that the spectrum of this GC
is clearly different from other GCs of the same colour in a sense
consistent with the derived younger age and higher metallicity,
i.e. similar Balmer line strengths and stronger metal lines (see
Fig. \ref{spectra}). It is clear from Fig. \ref{colour_comp} that this
effect is not the result of extinction. However, the proximity of this
GC to the galactic centre makes contamination by the background galaxy
a concern.  We therefore experimented with adding galaxy light back
into the GC spectrum and then subjecting the resultant spectrum to our
age/metallicity analysis. We found that when 50\% of the galaxy light
was recombined with the GC spectrum the derived age increased by
~0.2~dex (i.e to ~5~Gyr), while the metallicity fell by a similar
amount, resulting in a similar predicted colour. This is both a
relatively small change (for a relatively large amount of galaxy
contamination) \emph{and} is in the opposite sense to that required to
explain the young age by galaxy contamination. We therefore conclude
that background contamination in the spectroscopic analysis is
unlikely to be the cause of the observed young age of this GC, or the
discrepancy with its predicted colour.  The cause of the discrepancy
between observed and predicted colours therefore remains unknown.\\

\subsection{Stellar population parameters}
Having gained some confidence in our measured stellar parameters 
we now return our attention to the age and metallicity estimates. 

Our stellar population analysis identifies a single
young GC, with a derived age of 3.3~Gyr. This is similar to the
\emph{luminosity-weighted} age of 4.7~Gyr found for the galactic
centre by Proctor \& Sansom (2002). The central [Fe/H]=--0.03$\pm$0.09
and [E/Fe]=0.20$\pm$0.04 found in Proctor \& Sansom (2002) are also
similar to the values found for this young GC (--0.17$\pm$0.04 and
0.16$\pm$0.03 respectively; Fig. \ref{agez}). This
suggests the possibility that this GC formed in the same event that
fuelled the central star-burst.

We find the remaining 14 of 15 GCs to possess ages older than 10~Gyr
(Fig. \ref{agez}).  In five cases we find an age equal to the oldest
age modelled by Lee \& Worthey (2005).  It is apparent that the
scatter in GC age estimates is smaller than the error given by our
Monte-Carlo analysis (Section \ref{meas}). We take this to be a
combination of three effects; i) the error includes both random
\emph{and} systematic errors, ii) the scatter is slightly suppressed
by the GCs hitting the oldest age, iii) a slight over-estimation of
observational errors is also a possibility (see
Section \ref{meas}).

The 14 GCs found to be old span a broad range of metallicities
(Fig. \ref{agez}), similar to that observed in other spiral galaxy GC
systems (Burstein et al. 1984; Beasley et al. 2004; Schroder et
al. 2002; Larsen et al. 2002; Olsen et al. 2004). They also span a
similar range to the Milky Way GC system (de Angeli et al. 2005;
Fig. \ref{agez}).
 
\begin{table*}
\begin{centering}
\begin{tabular}[b]{|c|c|c|c|c|c|}
\hline 
ID   & Age &Log(age)     & [Fe/H] & [E/Fe]       & [Z/H]\\ 
& (Gyr)&(Gyr)&&&\\ 	 	  
\hline 		
gc03 &11.9   &   1.08(0.15)&  -1.40(0.13) &  0.21(0.08)&  -1.20(0.16) \\ 
gc04 &11.2   &   1.05(0.13)&  -0.99(0.17) &  0.09(0.08)&  -0.90(0.15) \\ 
gc06 &11.9   &   1.08(0.20)&  -1.57(0.33) &  0.21(0.14)&  -1.38(0.24) \\ 
gc07 &11.9   &   1.08(0.18)&  -1.63(0.26) &  0.24(0.11)&  -1.40(0.22) \\ 
gc08 &10.0   &   1.00(0.08)&  -1.95(0.09) &  0.24(0.15)&  -1.73(0.13) \\ 
gc09 &10.0   &   1.00(0.18)&  -0.72(0.17) &  0.05(0.07)&  -0.68(0.21) \\ 
gc10 &10.0   &   1.00(0.18)&   0.16(0.18) & -0.25(0.24)&  -0.08(0.24) \\ 
gc11 &3.3    &   0.53(0.02)&  -0.17(0.03) &  0.16(0.02)&  -0.03(0.03) \\ 
gc14 &10.0   &   1.00(0.07)&  -0.53(0.12) &  0.01(0.11)&  -0.53(0.08) \\ 
gc16 &11.2   &   1.05(0.07)&  -0.71(0.10) &  0.14(0.10)&  -0.58(0.12) \\ 
gc17 &11.2   &   1.05(0.14)&  -1.47(0.12) &  0.15(0.05)&  -1.33(0.09) \\ 
gc18 &10.0   &   1.00(0.10)&  -1.59(0.24) &  0.36(0.25)&  -1.25(0.16) \\ 
gc19 &11.9   &   1.08(0.15)&  -1.26(0.11) &  0.12(0.05)&  -1.15(0.12) \\ 
gc21 &11.2   &   1.05(0.04)&  -2.55(0.19) &  0.24(0.11)&  -2.33(0.18) \\ 
gc22 &11.9   &   1.08(0.26)&  -2.05(0.56) &  0.27(0.27)&  -1.80(0.52) \\ 
\hline
\end{tabular}
\caption{Derived values of age, [Fe/H], [E/Fe] and [Z/H] for 15 GCs.
 Errors (in brackets) represent statistical errors derived from 50
 Monte Carlo realisations of best fit data. All but one GC exhibit old
 ($>$10~Gyr) ages.}
\label{tab_agez}
\end{centering}
\end{table*}

Fig. \ref{agez} also shows a comparison of [E/Fe] values from our
study to the [$\alpha$/Fe] results of Pritzl et al. (2005).  We show
in Mendel et al. (2007) that, for Galactic GCs, there is good
agreement between [E/Fe] from Lick studies using Lee \& Worthey (2005)
models, and the [$\alpha$/Fe] results of Pritzl, Venn \& Irwin (2005).
The data suggest a slightly lower [E/Fe] in NGC~2683 than in the Milky
Way, but a larger, high signal-to-noise sample is required before we can 
draw any firm conclusions.\\

\subsection{Radial metallicity distribution}
\begin{figure}
\includegraphics[width=7cm,angle=-90]{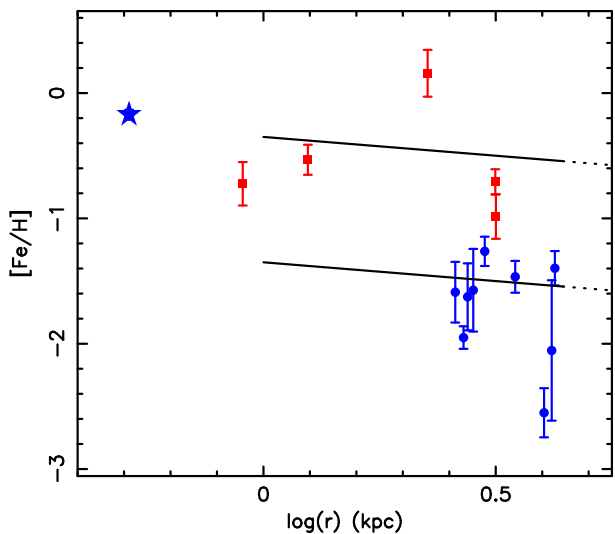}
\caption{[Fe/H] is plotted against projected distance
along the major axis. Symbols as Fig. \ref{agez}. Lines represent the
Harris (2000) trends for the average [Fe/H] with radius in Galactic
GCs when subdivided into red and blue subpopulations.}
\label{raddist}
\end{figure}

The final step in our analysis is to consider the radial distribution
in GC metallicities. To this end, a plot of [Fe/H] with radial
distance along the major axis is presented in Fig. \ref{raddist}.  

Radial trends in GC metallicity with galactocentric radius
are expected in a dissipative formation scenario. 
The GC systems of both M31 (Barmby et al. 2000) and Milky Way
(Armandroff, Da Costa \& Zinn 1992) have been found to exhibit little,
or no, overall radial metallicity gradient. However, Harris (2000)
shows that weak trends in metallicity with radius \emph{are} present 
when red and blue GC subpopulations are considered separately. More 
recently, Lee et al. (2007) showed that trends between metallicity 
and orbital parameters are present in the sub-sample of the Milky Way
population that excludes many blue GCs with extreme horizontal-branch
morphologies. Citing the extreme horizontal-branch GCs as probable
accreted (and stripped) dwarf galaxies, Lee et al. (2007) conclude
that the `normal' GCs show clear signs of dissipational collapse.

We find no evidence for trends with azimuthal distance (perpendicular
to the galaxy rotation plane) in NGC~2683, although we note the
extremely limited range of our data.  Our data \emph{do}, on the other
hand, suggest a trend of decreasing GC [Fe/H] with increasing distance
along the major axis of NGC~2683 with logarithmic slope of --1.7
(Fig. \ref{raddist}).  However, the data for both red and blue GCs of
NGC~2683 are also consistent with the Harris (2000) trends for the
average [Fe/H] with radius in Galactic GCs. In
the Milky Way, there is no significant trend in the GC system as a
whole, but individually both red and blue sub-populations show weak
trends with radius of logarithmic slope --0.3, albeit with
considerable scatter. We also note that our sample for NGC~2683
contains no blue GCs within $\sim$2~kpc, and no red GCs beyond
$\sim$3~kpc, while the photometry of Forde et al. (2007) clearly shows
that both red and blue GCs are present throughout the radius range
covered by our data.  Therefore, we suspect that the apparent trend is
the result of the lack of observations of red GCs at large radii and
blue GCs at small radii, and is consequently simply a sampling issue.
A definitive description of this aspect of the GC system of NGC~2683
must, however, await a more extensive study.
 
\section{Conclusions}
\label{disc}
We have analysed the recession velocities and stellar populations of a
small sample of GCs in the spiral galaxy NGC~2683 and
compared the results with the Galactic GC system.

Our stellar population analysis identified one GC, located near the
centre of NGC~2683, with the relatively young age
of 3.3$\pm$0.5~Gyr. The age, metallicity and [E/Fe] of this young GC
appear remarkably similar to the values found by Proctor \& Sansom (2002)
for the centre of NGC~2683 itself. This result therefore suggests the
possibility that this GC and the recent burst in the central
regions were formed at the same time, from the same gas supply, and
provide further evidence for a star-formation event in NGC~2683 about
$\sim$3~Gyr ago.

The stellar population parameters of the 14 \emph{old} globular
clusters in our sample show many similarities to the old globular
clusters of the Milky Way. The metallicity distribution spans a
similar range to those found in studies of the Milky Way and other
spiral galaxy systems, i.e from $\sim$--2.5 to 0.0~dex. 
The data for NGC~2683 GCs are also consistent with
the trends in [Fe/H] with radius observed in red and blue Galactic GC
subpopulations.\\

\section{Acknowledgements}
We thank Soeren Larsen for help preparing the slit mask and Kieran Forde 
for providing information prior to publication. We also thank Lee Spitler
for assistance with the photometric analysis.  
Part of this research was funded by NSF grant AST-02-06139 The data presented
herein were obtained at the W.M. Keck Observatory, which is operated
as a scientific partnership among the California Institute of
Technology, the University of California and the National Aeronautics
and Space Administration.  The Observatory was made possible by the
generous financial support of the W.M. Keck Foundation. This research
has made use of the NASA/IPAC Extragalactic Database (NED), which is
operated by the Jet Propulsion Laboratory, Caltech, under contract
with the National Aeronautics and Space Administration. We also 
thank the Australian Research Council for funding that supported this work.\\

\section{References}
Armandroff T.E., Da Costa G.S., Zinn R., 1992, AJ, 104, 164\\
Ashman K.M., Zepf, S.E., 1992, ApJ, 384, 50\\
Barbon R., Capaccioli M., 1975, A\&A, 42, 221\\
Barmby P., Huchra J.P., Brodie J.P., Forbes D.A., Schroder L.L., Grillmair C.J. 2000, AJ, 119, 727\\
Beasley M.A., Brodie J.P., Strader J., Forbes D.A., Proctor R.N., Barmby P., Huchra J.P., 2004, AJ, 128, 1623\\
Bedregal A.G., Aragón-Salamanca A., Merrifield M.R., 2006, MNRAS, 373, 1125\\
Broeils A.H., van Woerden H., 1994, A\&AS, 107, 129\\
Bruzual A.G., Charlot S., 2003, MNRAS, 344, 1000\\
Burstein D., Faber S.M., Gaskell C.M., Krumm N., 1984, ApJ, 287, 586\\
Casertano S., van Gorkom J.H., 1991, AJ, 101, 1231\\
Chandar R., Puzia T.H., Sarajedini A., Goudfrooij  P., 2006, ApJ, 646L, 107\\
Courteau S., van den Bergh S., 1999, AJ, 118, 337\\
De Angeli F., Piotto G., Cassisi S., Busso G., Recio-Blanco A., Salaris M., Aparicio A., Rosenberg A., 2005, AJ, 130, 116\\
Eggen O.J., Lynden-Bell D., Sandage A.R., 1962, ApJ, 136, 748\\
Forbes D.A., Beasley M.A., Bekki K., Brodie J.P., Strader J.,  2003, Science, 301, 1217\\
Forbes D.A., Brodie J.P., Larsen S.S., 2001, ApJ, 556 , 83\\
Forbes D.A., Strader J., Brodie J.P., 2004, AJ, 127, 3394\\
Forde et al., 2007, in preparation\\
Gilmore G., King I., van der Kruit P., 1989, Proceedings of the 19th Advanced Course of the Swiss Society of Astronomy and Astrophysics (SSAA), Saas-Fee, Leysin, Vaud, Switzerland, 13-18 March, 1989, Geneva: Observatory, 1989, edited by Buser, Roland, p334\\
Goudfrooij P., Strader J., Brenneman L., Kissler-Patig M., Minniti D., Huizinga J.E., 2003, MNRAS, 343, 665\\
Harris W.E., 2000, ``Star Clusters'', in ``28th Saa-Fee Advanced Course for Astrophysics and Astronomy''\\
Houdashelt M.L., Trager S.C., Worthey G., Bell R.A., 2002, Elemental Abundances in Old Stars and Damped Lyman-α Systems, 25th meeting of the IAU, Joint Discussion 15, 22 July 2003, Sydney, Australia\\
Ibata R.A., Gilmore G., Irwin M.J., 1995, MNRAS, 277, 781\\
Jensen J.B., Tonry J.L., Barris B.J., Thompson R.I., Liu M.C., Rieke M.J., Ajhar E.A., Blakeslee J.P., 2003, ApJ, 583, 712\\ 
Kent S.M., 1985, ApJS, 59, 115\\
Kissler-Patig M., Ashman K.M., Zepf S.E., Freeman K.C., 1999, AJ, 118, 197\\
Lanfranchi G.A., Matteucci F., 2004, MNRAS, 351, 1338\\  
Larsen S.S., Forbes D.A., Brodie J.P., 2001, MNRAS, 327, 1116\\
Larsen S.S., Brodie J.P., Beasley M.A., Forbes D.A., 2002, AJ, 124, 828\\
Lee H., Worthey G.,  2005, ApJS, 160, 176\\
Lee Y-K., Gim H.B., Casetti-Dinescu D.I., ApJ, 661, L52\\
Mackey A.D., Gilmore G.F., 2004, MNRAS, 355, 504\\
Martin N.F., Ibata R.A., Bellazzini M., Irwin M.J., Lewis G.F., Dehnen W., 2004, MNRAS, 348, 12\\
Mendel J.T., Proctor R.N., Forbes D.A., 2007, MNRAS, 379, 1618\\
Merrifield M.R., Kuijken K., 1999, A\&A, 345, 47\\
Oke J.B., Cohen J.G., Carr M., Cromer J., Dingizian A., Harris F.H., Labrecque S., Lucinio R., Schaal W., Epps H., Miller J., 1995, PASP, 107, 375\\
Olsen K.A.G., Miller B.W., Suntzeff N.B., Schommer R.A., Bright J., 2004, AJ, 127, 2674\\
Pierce M., Brodie J.P., Forbes D.A., Beasley M.A., Proctor R.N., Strader J., 2005, MNRAS, 358., 419\\
Pierce M., Bridges T., Forbes D.A., Proctor R.N., Beasley M.A., Gebhardt K., Faifer F.R., Forte J.C., Zepf S.E., Sharples R., Hanes D.A., 2006, MNRAS, 368. 325\\
Pritzl B.J., Venn K.A., Irwin M., 2005, AJ, 130, 2140\\
Proctor R.N., Sansom A.E., 2002, MNRAS, 333, 517\\
Proctor R.N., Forbes D.A., Beasley M.A.,  2004a, MNRAS, 355, 1327\\
Proctor R.N., Forbes D.A., Hau G.K.T., Beasley M.A., De Silva G.M., Contreras R., Terlevich A.I., 2004b, MNRAS, 349, 1381\\
Proctor R.N., Forbes D.A., Forestell A., Gebhardt K., 2005, MNRAS, 362, 857\\
Rhode K.L., Zepf S.E., Kundu A., Larner A.N., astro-ph/0708.1166\\
Schroder L.L., Brodie J.P., Kissler-Patig M., Huchra J,P., Phillips A.C., 2002, AJ, 123, 2473\\
Searle L., Zinn R.,  1978, ApJ, 225, 357\\
Strader J., Smith G., 2007, ApJ, submitted\\
Thomas D., Maraston C., Bender R., 2003, MNRAS, 339, 897\\
Tonry J.L., Dressler A., Blakeslee J.P., Ajhar E. A., Fletcher A.B., Luppino G.A., Metzger M.R., Moore C.B., 2001, ApJ, 546, 681\\
Trager S.C., Worthey G., Faber S.M., Burstein D., Gonzarlez J.J., 1998, ApJS, 116, 1\\
van den Bergh S., 1999, A\&ARv, 9, 273\\
Worthey G., Ottaviani D.L., 1997, ApJS, 111, 377\\

\begin{appendix}

\section{Lick indices}

\begin{table*}
\tiny
\begin{centering}
\begin{tabular}[b]{|c|c|c|c|c|c|c|c|c|c|c|c|c|c|c|}  
\hline
  GC   &  Ca4227 & G4300 &H$\gamma_A$&H$\gamma_F$&Fe4383&  Ca4455&  Fe4531&  C4668 &H$\beta$&  Fe5015 &   Mgb   &  Fe5270&  Fe5335&  Fe5406  \\
&$\AA$&$\AA$&$\AA$&$\AA$&$\AA$&$\AA$&$\AA$&$\AA$&$\AA$&$\AA$&$\AA$&$\AA$&$\AA$&$\AA$\\
\hline
gc03   &  0.613  & 2.496 &(-0.773)   & 1.357    & 2.356 &  0.878 &  1.667 &  0.856 &  2.407 & (3.319) &  (0.789)&  1.459 &  1.354 &  0.473\\
       &  0.158  & 0.308 & (0.284)   & 0.150    & 0.382 &  0.195 &  0.285 &  0.413 &  0.202 & (0.396) &  (0.195)&  0.230 &  0.278 &  0.203\\
gc04   &  0.826  &(2.314)& -1.802    &(0.650)   & 4.043 & (2.170)&  3.129 &  3.086 &  0.849 &  3.717  &   1.704 &  1.537 &  1.434 &  1.024\\
       &  0.252  &(0.497)&  0.484    &(0.283)   & 0.644 & (0.335)&  0.513 &  0.776 &  0.347 &  0.706  &   0.327 &  0.388 &  0.457 &  0.341\\
gc05   &  0.491  & 4.129 & -1.598    & 1.097    &(0.016)&  1.410 &  2.178 &  3.120 &  2.710 &  5.260  &  (3.361)&  2.823 &  2.081 & (2.848)\\
       &  0.297  & 0.515 &  0.543    & 0.316    &(0.800)&  0.400 &  0.596 &  0.898 &  0.375 &  0.847  &  (0.406)&  0.481 &  0.558 & (0.388)\\
gc06   &  0.558  & 1.760 &  1.102    & 1.957    &(-0.892)& 0.026 & (4.640)& -0.698 & (2.973)&    --   &  (2.502)& (2.594)& (3.365)&  0.975\\
       &  0.272  & 0.498 &  0.457    & 0.268    &(0.756)&  0.368 & (0.505)&  0.823 & (0.334)&    --   &  (0.349)& (0.401)& (0.459)&  0.360\\
gc07   &  0.674  & 1.421 & (2.184)   & 2.137    & 1.481 & (1.445)&  1.969 &  1.189 & (1.944)&(-0.967) &   1.359 &  0.712 & (2.443)& (1.234)\\
       &  0.259  & 0.493 & (0.455)   & 0.273    & 0.706 & (0.354)&  0.567 &  0.853 & (0.357)& (0.811) &   0.367 &  0.445 & (0.508)& (0.392)\\
gc08   &  0.270  &(2.277)&  1.095    &(2.398)   & 1.525 &  0.543 &  1.005 &(-0.971)&  2.817 &  1.963  &  (0.889)&  0.527 &(-0.296)&  0.478\\
       &  0.139  &(0.277)&  0.245    &(0.121)   & 0.335 &  0.170 &  0.249 & (0.354)&  0.179 &  0.341  &  (0.159)&  0.196 & (0.239)&  0.168\\
gc09   &  0.987  & 4.497 &(-1.567)   &(1.870)   & 4.216 &  1.480 &  2.695 & (3.496)& (3.054)&  4.700  &   2.082 &  1.809 &  2.001 &  1.139\\
       &  0.146  & 0.288 &( 0.274)   &(0.138)   & 0.347 &  0.176 &  0.261 & (0.375)& (0.192)&  0.374  &   0.177 &  0.213 &  0.250 &  0.183\\
gc10   & (2.085) &(4.412)& -6.486    &(-0.820)  & 5.984 &  1.888 &  3.711 &(-0.545)& (3.132)&  4.335  &  (5.628)&  3.803 &  3.382 &(-1.105)\\
       & (0.266) &(0.571)&  0.672    &(0.397)   & 0.806 &  0.417 &  0.621 & (1.018)& (0.378)&  0.908  &  (0.402)&  0.496 &  0.589 & (0.508)\\
gc11   & (1.282) & 4.757 & -3.718    & 0.044    & 4.424 &  1.341 &  2.790 &  4.384 &  2.302 & (4.093) &   3.153 &  2.577 &    --  &  1.314\\
       & (0.101) & 0.226 &  0.193    & 0.081    & 0.221 &  0.110 &  0.143 &  0.160 &  0.136 & (0.192) &   0.080 &  0.109 &    --  &  0.092\\
gc14   & (2.066) &(8.325)&(-6.007)   &-0.872    & 3.526 &  1.497 &  2.655 &  2.910 & (2.760)& (7.069) &  (6.047)& (3.092)&  2.487 & (2.594)\\
       & (0.152) &(0.310)& (0.349)   & 0.198    & 0.427 &  0.219 &  0.314 &  0.454 & (0.214)& (0.407) &  (0.181)& (0.232)&  0.278 & (0.193)\\
gc16   & (1.335) & 4.365 & -4.644    &-0.772    & 4.343 &  1.316 &  1.977 & (0.293)&  1.886 & (5.444) &  (1.715)&  2.616 &  1.721 &  1.147\\
       & (0.177) & 0.350 &  0.364    & 0.211    & 0.454 &  0.236 &  0.354 & (0.530)&  0.237 & (0.463) &  (0.227)&  0.255 &  0.311 &  0.226\\
gc17   &  0.812  & 2.337 & -0.019    &(1.640)   & 1.692 & (0.951)&  1.631 & (0.399)& (2.379)&  2.857  &   1.007 & (1.963)& (1.882)&  0.660\\
       &  0.134  & 0.277 &  0.247    &(0.124)   & 0.331 & (0.167)&  0.241 & (0.340)& (0.177)&  0.324  &   0.151 & (0.178)& (0.220)&  0.159\\
gc18   &  0.602  &(4.593)& -1.077    & 0.433    & 1.818 &  1.729 &(-0.295)&(-3.244)& (3.752)& (6.285) &  (6.797)&  1.817 & (2.002)&  0.362\\
       &  0.191  &(0.358)&  0.379    & 0.223    & 0.553 &  0.259 & (0.452)& (0.660)& (0.273)& (0.597) &  (0.260)&  0.364 & (0.429)&  0.340\\
gc19   &  0.347  & 2.742 & -0.704    & 1.413    & 2.508 &  0.893 &  2.543 & (1.323)&  1.885 &  3.133  &   1.415 & (2.099)&  1.694 &  0.657\\
       &  0.168  & 0.319 &  0.299    & 0.160    & 0.404 &  0.208 &  0.307 & (0.454)&  0.214 &  0.426  &   0.205 & (0.241)&  0.292 &  0.217\\
gc21   &  0.414  & 0.368 &  2.799    & 3.054    & 1.560 &  0.614 &  0.488 &(-3.468)&  3.386 &  1.508  &  (1.695)& -1.085 &  1.114 &  0.011\\
       &  0.250  & 0.470 &  0.406    & 0.228    & 0.629 &  0.326 &  0.533 & (0.810)&  0.316 &  0.732  &  (0.347)&  0.451 &  0.510 &  0.392\\
gc22   &  0.554  & 1.470 &  2.364    &(3.414)   & 1.936 &  1.382 & -1.338 &   --   &    --  &(-0.760) &     --  &(-2.791)&(-3.424)& (3.388)\\
       &  0.350  & 0.795 &  0.672    &(0.393)   & 1.015 &  0.570 &  0.967 &   --   &    --  & (1.420) &     --  & (0.840)& (1.076)& (0.642)\\
\hline
\end{tabular}
\caption{Fully corrected Lick indices are shown. Indices in brackets were excluded from the fitting procedure used to derive ages and metallicities.}
\label{indices}
\end{centering}
\normalsize
\end{table*}

\end{appendix}

\end{document}